\documentclass[aip,authoryear,floats,showpacs,amssymb,tightenlines]{revtex4}

\usepackage{pifont}
\usepackage{amsmath}
\usepackage{amsfonts}
\usepackage{amscd}
\usepackage{epsfig}
\usepackage{amssymb}
\usepackage{tabularx}
\usepackage[english]{babel}
\usepackage[utf8]{inputenc}
\usepackage{ae,aecompl}
\usepackage{color}

\newcommand{\unit}[1]{\ensuremath{\, \mathrm{#1}}}

\begin{document}

\title{A simple mechanism for the anti-glitch observed in magnetar 1E 2259+586}

\author{Federico Garc\'{\i}a}
 \email{fgarcia@iar-conicet.gov.ar, fgarcia@fcaglp.unlp.edu.ar}
\affiliation {Instituto Argentino de Radioastronom\'{\i}a, CCT La Plata - CONICET, C.C. 5 (1984) Villa Elisa, Buenos Aires, Argentina \\ Facultad de Ciencias Astron\'omicas y Geof\'{\i}sicas, Universidad Nacional de La Plata.
 Paseo del Bosque S/N 1900. La Plata, Argentina}
\author{Ignacio F. Ranea-Sandoval}
  \email{iranea@fcaglp.unlp.edu.ar}
\affiliation{Grupo de Gravitaci\'on, Astrof\'{\i}sica y Cosmolog\'{\i}a, Facultad de Ciencias Astron\'omicas y Geof\'{\i}sicas, Universidad Nacional de La Plata.
 Paseo del Bosque S/N 1900. La Plata, Argentina}

\begin{abstract}
In this short communication we present a simple internal mechanism that accounts for the recently observed anti-glitch in magnetar 1E 2259+586.

We propose that the decay of an internal toroidal magnetic field component would de-stabilize an originally stable prolate star configuration. Then, the subsequent rearrangement of the stellar structure would give rise to a ``more spherical'' configuration, resulting in a sudden spin-down of the whole star.

We present here some order of magnitude calculation to give confidence to this scenario, using the simplest analytical stellar model and let more detailed calculations for a more technical future paper.
\end{abstract}


\maketitle

\tableofcontents

\section{Preliminaries}

Magnetars are neutron stars powered by their strong internal magnetic fields \citep{duncan1992}. The detailed study of the temporal behaviour of magnetar's emission can be used to constrain both their external dipolar magnetic field and internal structure \citep{kouve1998,glitches}. Despite the fact that all neutron stars suffer a long term spin-down, due to their dipole magnetic field decay, many sudden spin-up, known as glitches in the literature, have been observed in pulsars and magnetars \citep{glitches1,glitches2}. Recently, clear evidence of the first sudden spin-down detection was observed in magnetar 1E 2259+586 \citep{antiglinch}, a so-called ``anti-glitch''. There, the authors propose two different interpretations for the observational evidence: (i) an anti-glitch event followed by a normal glitch (after $\sim$100 days) and (ii) a sequence of two anti-glitches (separated by $\sim$50 days). Spin-down events were also observed in magnetars SGR 1900+14 \citep{sgr1900} and 4U 0142+61 \citep{4u}, and in the high magnetic field pulsar PSR J1846--0258 \citep{psrj}, but none of them is considered as an anti-glitch, because of their long timescales ($17-127$~days). 

The Anomalous X-ray Pulsar 1E 2259+586, with a $\sim 7$~s period, is a magnetar with a characteristic age $\sim 10^4$~yr and a spin-inferred surface dipolar magnetic field of $B_{d} \sim 5.9 \times 10^{13}$~G, which poses a minimum value for the internal magnetic field strength. 1E 2259+586 has been monitored with {\it Rossi X-ray Timing Explorer} and {\it Swift X-ray Telescope} over almost the last two decades, showing a stable spin-down rate, with the exception of two spin-up glitches in 2002 \citep{kaspi2003} and 2007 \citep{idem2012}, a timing event in 2009 \citep{idem2012}, and this anti-glitch in 2012.

From \cite{antiglinch}, it follows that the observational fact that has to be explained by any model is a total change in frequency in the complete $\sim 100$ day event was $\sim -5 \times 10^{-7}$~Hz, that can be interpreted as two instantaneous changes in spin frequency $(|\Delta \nu / \nu| \gtrsim  10^{-7})$. In order to adjust the timing data from {\it Swift}, the authors propose two models: (i) an anti-glitch in which $\Delta \nu / \nu = -3.1(4) \times 10^{-7}$ followed by a spin-up event, of amplitude $\Delta \nu / \nu = 2.6(5) \times 10^{-7}$; (ii) an anti-glitch in which $\Delta \nu / \nu = -6.3(7) \times 10^{-7}$ followed by a second anti-glitch in which $\Delta \nu / \nu = -4.8(5) \times 10^{-7}$. Based on a bayesian analysis, model (ii) is favoured \citep{hu2013}.

Several explanations for this anti-glitch event have been proposed, based both on external \cite{antiglinch-explanations} and internal \citep{duncan2013} origins. Despite searches in radio and X-ray wavelengths, no surrounding afterglow was detected \citep{antiglinch}, arguing against a sudden particle outflow or wind-driven scenario. In this sense, a most promising approach seems to be a an internal rearrangement of the star. With this short communication we aim to provide a simple mechanism where the sudden spin-down of the star occurs while the stellar structure reaccommodates after a substantial decay of the internal toroidal magnetic field component. A similar scenario was also suggested to account for the SGR 1900+14 event \citep{ioka}.

In \citep{cutler.2002,mag-deform}, the authors developed a formalism to model magnetically deformed neutron stars. In their work, deformations are calculated for a uniform density star with a mixed poloidal-toroidal magnetic field configuration, rotating with angular velocity, which, in the case of magnetars, as 1E 2259+586, is completely irrelevant, because of their long periods. The main result of their work is that while for strong poloidal magnetic fields, as for rapid rotation, the stars tend to oblate to keep mechanically stable, if internal toroidal fields dominate the magnetic field configuration, prolate stars are favoured.

For volume preserving $l=2$ mode, and mixed toroidal-poloidal magnetic field, the quadrupolar distortion for incompressible stars of uniform density is given by:

\begin{equation} \label{epsilon}
\epsilon = \frac{I_{ zz} - I_{xx}}{I_{zz}} = - \frac{25R^4}{24G_NM^2}\left(\left<B_t^2\right>-\frac{21}{10}\left<B_p^2\right>\right),
\end{equation}

\noindent where $\left<B_t^2\right>$ is the mean value of the square of the toroidal magnetic field strength, $\left<B_p^2\right>$ is the mean value of the poloidal component, $R$ the radius of the undeformed star, $G_N$ the gravitational constant and $M$ the mass of the neutron star.

Despite that a purely toroidal magnetic field is unstable \cite{instab-toro, instab-polo1, instab-polo2, stable}, an additional poloidal component with energy $E_p/E_t = B_p^2/B_t^2 \sim 1-5\%$ stabilizes the magnetic field configuration \cite{reisenegger2013}, allowing us to neglect the poloidal contribution to Eq. \ref{epsilon} as $\left<B_p^2\right> \ll \left<B_t^2\right>$.

\section{The model}

Without entering in detailed calculations, based on Eq. (\ref{epsilon}) it is clear that the net effect of the decay of the toroidal magnetic field, which is expected due to the combination of Ohmic and Hall effects in the neutron star interior, as it is shown in numerical simulations \citep[see, fon instance,][]{pons2009,vigano2013}, is to make an initially stable prolate configuration unstable. The ``more spherical'' configuration has greater moment of inertia respect to the rotation axis, $z$ in this case, and thus, in the abscense of an external torque, due to the conservation of the angular momentum, this could easily account for the sudden spin-down observed.


The theoretical picture that we want to explode is the following: given a certain ``initial'' mostly toroidal magnetic field strength, $\left<B_t^i\right>$, the neutron star crust crystallizes in a prolate equilibrium configuration. Then, as the magnetic field decays this prolate configuration becomes unstable up to a point in which a sudden change in the shape of the star occurs. After this, the stellar structure achieves a new stable and less prolate configuration associated to $\left<B_t^f\right>$. Based on Eq. \ref{epsilon}, a change in the effective toroidal magnetic field component will produce a change in the moments of inertia that would lead to a change in the rotation frequency of the star. If the magnetic field axis is aligned with the rotation axis of the star, the relative change in frequency will be given by:

\begin{equation} \label{deltanu-o-nu}
\frac{\Delta \nu}{\nu ^{i}} \equiv \frac{\nu ^{f} - \nu ^{i}}{\nu ^{i}} = \frac{I_{zz}^i}{I_{zz}^f} -1 = \left(\frac{R_e^i}{R_e^f}\right)^2 -1,
\end{equation}

\noindent where $R_e$ is the equatorial radius of the star. Actually, we can analytically calculate $R_e$ as a function of $\left<B_t^2\right>$ using Eq.~(\ref{epsilon}), from where we obtain:

\begin{equation}
R_e = R (1 - 2\epsilon)^{-1/6}.
\end{equation}

\noindent Thus, putting all together we can write (Eq. \ref{deltanu-o-nu}) as as a function of the quadrupolar distortion, resulting:

\begin{equation} \label{plot}
\frac{\Delta \nu}{\nu} = \frac{(1-2\epsilon ^f)^{1/3}}{(1-2\epsilon ^i)^{1/3}} - 1 \sim \frac{2}{3}(e^i - e^f).
\end{equation}

\noindent where, in the last approximation, we used that, in the case of our interest, both $\left|e^{i,f}\right| \sim 10^{-6} \ll 1$.

\section{Results}

Following the model presented in Section II, for a constant density neutron star, which is deformed to a prolate shape by the presence a mostly toroidal magnetic field \citep{cutler.2002}, we calculate the change in the magnetic field strength needed to account for the $\Delta \nu / \nu$ observed in 1E 2259+586. 

\begin{figure}
 \centering
 \includegraphics[width=10 cm]{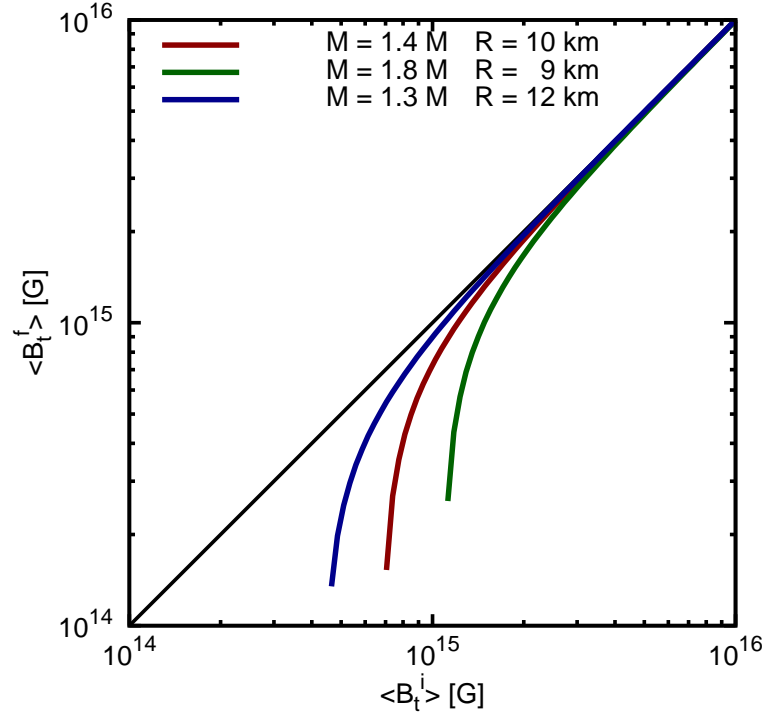}
 \caption{Physical solutions to Eq. (\ref{plot}) for $\Delta \nu /\nu = - 6.3 \times 10^{-7}$ as function of $\left<B_t^i\right>$, for three different neutron star configurations. In solid black we plot the identity function as a reference.}
 \label{fig1}
\end{figure}

For a ``typical'' neutron star of radius $R = 10 \unit{km}$ and mass $M = 1.4 \unit{M_\odot}$, with a mean toroidal magnetic field of $\left<B_t\right>= 2 \times 10^{15}$~G, which corresponds to a maximum value $B_M \gtrsim 10^{16}$~G \citep{reisenegger2013}, we estimate that a decay in the magnetic field of about $\sim 10\%$ can be responsible for the observed spin-down. These qualitative result is almost insensitive to other acceptable values for mass, radius and magnetic field strength (see Figure \ref{fig1}). 

In Figure (\ref{fig1}) we plot the physical solutions, $\Delta \left<B_t\right> < 0$, to Eq. (\ref{plot}), as a function of the initial toroidal magnetic field strength $\left<B_t^i\right>$, for three different neutron star configurations, assuming a jump in frequency $\Delta \nu / \nu = -6.3 \times 10^{-7}$ equal to the first of the two events of model (ii). In black we plot the identity as a reference. 

For each of the adopted neutron star configurations, we find that a minimum value for $\left<B_t^i\right>$ is needed in order to have a solution to Eq. (\ref{plot}) for the observed $\Delta \nu / \nu$. This critical value is, in any case, several times $10^{14}$~G, which avoids the occurrence of anti-glitches in normal pulsars, only allowing this phenomena to occur in strongly magnetized neutron stars, i.e. magnetars. This result explains why despite that many pulsars have been thoroughly monitored for several decades, no sudden spin-down event of this kind has been detected at all.



\begin{figure}
 \centering
 \includegraphics[width=10 cm]{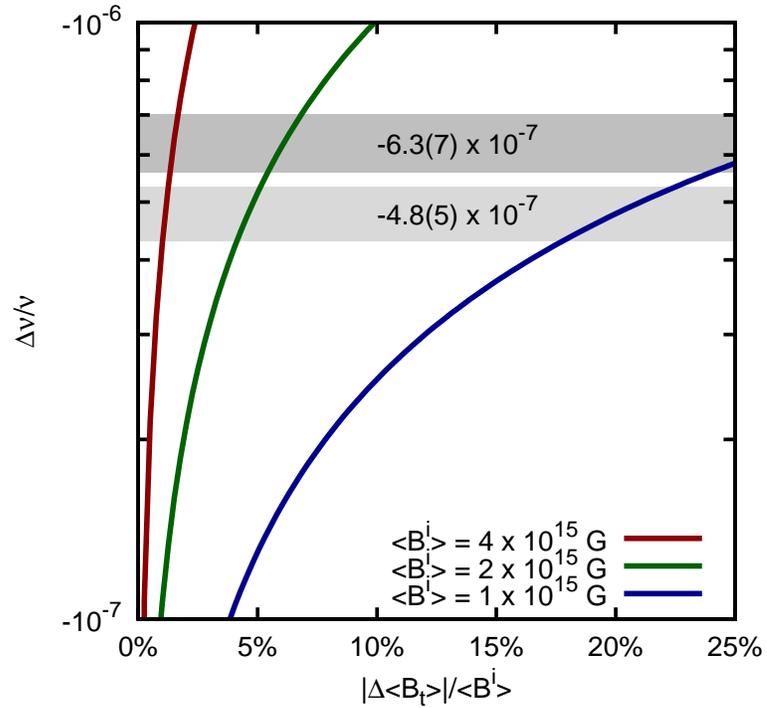}
 \caption{Calculated $\Delta \nu / \nu$ values from Eq. (\ref{plot}) as a function of the change $\left|\Delta \left<B_t\right>\right| / \left<B_t^i\right>$ in the toroidal magnetic flied strength. As a reference the values corresponding to the anti-glitch/anti-glitch pair are presented in horizontal lines with error bars considered.}
 \label{fig3}
\end{figure}

In Figure (\ref{fig3}) we present $\Delta \nu / \nu$ from Eq. (\ref{plot}) as a function of the change $\left|\Delta \left<B_t\right>\right| / \left<B_t^i\right>$ in the toroidal magnetic flied strength. The horizontal lines account for the values corresponding to model (ii), and the shaded rectangles for the error bars.

\section{Conclusions}

A very simple model to explain the anti-glitch observed in magnetar 1E 2259+586 is presented here. We propose that a natural decay of the toroidal magnetic field component, of approximately $\sim 10\%$, from an initial $\left<B_t^i\right> \gtrsim 10^{15}$~G, would be enough to de-stabilize an originally prolate configuration for the stellar structure, reducing it to a ``more spherical'' one. As a result, a sudden change in the moments of inertia of the star produces a net spin-down, as the one observed in \cite{antiglinch}. Detailed studies of the magnetic field evolution in neutron stars show that a magnetic field decay of $\sim 10\%$ as the one needed in our model is easily achieved after $t \lesssim 10^4$~yr for a magnetar like 1E 2259+586 \citep{vigano2013}. 

It is also worth to notice that gravitational waves potentially observed by advanced LIGO, should be emitted under this scenario due to oscillations induced by the changes in the stellar structure.

More involved calculations using more realistic neutron star configurations are being carried away \cite{con-fede-prep} to refine these results.


\section*{Acknowledgements}

We are thankful to H\'ector Vucetich, Deborah N. Aguilera for useful discussions, encouragement and feedback. FG is a Fellow of CONICET. IFRS is a Fellow of CONICET and acknowledges support by UNLP.

\end{document}